\documentclass[journal,onecolumn, draftcls, 12pt]{IEEEtran} 

\usepackage[utf8]{inputenc}
\usepackage[T1]{fontenc}
\usepackage[hidelinks]{hyperref}
\usepackage{url}
\usepackage{ifthen}
\usepackage{cite}
\usepackage[cmex10]{amsmath} 
\usepackage{amsfonts,amssymb}
\usepackage{color}
\usepackage{ifthen}
\usepackage[arxiv]{optional} 

\begin{document}
\title{Decentralized sequential active hypothesis testing and the MAC feedback
capacity}

 \author{%
   \IEEEauthorblockN{Achilleas Anastasopoulos and Sandeep Pradhan}
   \IEEEauthorblockA{University of Michigan\\
                     Ann Arbor, MI 48109, USA\\
                     Email: \texttt{\{anastas,pradhanv\}@umich.edu}}
 }

\maketitle

\begin{abstract}
We consider the problem of decentralized sequential active hypothesis testing (DSAHT), where two transmitting agents, each possessing a private message, are actively helping a third agent--and each other--to learn the message pair over a discrete memoryless multiple access channel (DM-MAC). The third agent (receiver) observes the noisy channel output, which is also available to the transmitting agents via noiseless feedback. We formulate this problem as a decentralized dynamic team, show that optimal transmission policies have a time-invariant domain, and characterize the solution through a dynamic program.
Several alternative formulations are discussed involving time-homogenous cost functions and/or variable-length codes, resulting in solutions described through fixed-point, Bellman-type equations.

Subsequently, we make connections with the problem of simplifying the multi-letter capacity expressions for the noiseless feedback capacity of the DM-MAC. We show that restricting attention to distributions induced by optimal transmission schemes for the DSAHT problem, without loss of optimality, transforms the capacity expression, so that it can be thought of as the average reward received by an appropriately defined stochastic dynamical system with time-invariant state space.
\end{abstract}

\optv{2col}{
\textit{A full version of this paper is accessible at:}
\url{https://arxiv.org/abs/2001.03807}
}


\def\cE{\mathcal{E}}
\def\cX{\mathcal{X}}
\def\cY{\mathcal{Y}}
\def\cZ{\mathcal{Z}}
\def\cW{\mathcal{W}}
\def\cP{\mathcal{P}}
\def\cU{\mathcal{U}}
\def\cV{\mathcal{V}}
\def\cR{\mathcal{R}}
\def\cC{\mathcal{C}}
\def\cS{\mathcal{S}}
\def\cF{\mathcal{F}}
\def\cG{\mathcal{G}}
\def\cB{\mathcal{B}}

\def\tw{\tilde{w}}

\newtheorem{lemma}{Lemma}
\newtheorem{fact}{Fact}
\newtheorem{theorem}{Theorem}

\newcommand{\ve}[1]{\underline{#1}}
\newcommand{\eqdef}{\stackrel{\scriptscriptstyle \triangle}{=}}
\newcommand{\mdef}{\stackrel{\text{\tiny def}}{=}}
\newcommand{\E}{\mathbb{E}}
\def\Real{\mathbb{R}}
\def\P{\mathbb{P}}

\vspace*{-0.15cm}
\section{Introduction}

Active hypothesis testing refers to the problem where an agent is adaptively selecting the most informative sensing action, from a set of available ones, in order to obtain information about an underlying phenomenon of interest (hypothesis). The term ``active'' emphasizes the fact that the agent can exert some control over the
sensing action.
This problem was originally introduced by Blackwell~\cite{Bl53} in its single-shot version.
The ``sequential'' aspect of this problem refers to the setting where sensing decisions are performed at each time instance based on the available information and state of knowledge of the decision agent, i.e., in a closed-loop fashion. This problem generalizes the classical sequential hypothesis testing~\cite{WaWo48} and has been studied originally by~\cite{Ch59}.

Decentralized sequential active hypothesis testing (DSAHT) refers to a setting where multiple agents, each with some partial information about the underlying phenomenon of interest, are actively collaborating in order to obtain information about the said phenomenon.
Transmission of information over a multiple access channel (MAC) with feedback can be thought of as an instance of a DSAHT problem. Indeed in this setting, two agents (transmitters), each possessing a private message, are actively helping a third agent (receiver) to learn the message pair by transmitting symbols to the common medium modeled as a MAC. The third agent (receiver) observes the noisy channel output, which is also available to the transmitting agents via noiseless feedback, giving rise to a sequential process.
The decentralized Wald (non-active) problem has been studied in~\cite{TeHo87}, and more recently, a more general setting was considered in~\cite{NaTe11b}. A real-time communication system with two encoders communicating with a single receiver over separate noisy channels without feedback was considered in~\cite{NaTe11}.

In the first part of this paper, we formulate the DSAHT over the MAC as a decentralized dynamic team problem.
We show that optimal encoders are not required to depend on the entire feedback history, but have a time-invariant domain. Specifically they only depend on their private message and an appropriately defined posterior belief on the message pair from the viewpoint of the receiver. This result is both intuitive and satisfying as it generalizes the optimal encoding schemes for point-to-point channels~\cite{Bu76,NaJa13}. Furthermore, we show that the optimal encoders are characterized through a dynamic program. Several alternative formulations are discussed involving time-homogenous cost functions and/or variable-length codes, resulting in solutions described through fixed-point, Bellman-type equations.

In the second part of this paper we discuss how the above results can shed light on the problem of characterizing the MAC feedback capacity. A multi-letter capacity expression for DM-MAC with noiseless feedback has been established in~\cite{Kr98} and restated in~\cite{Sa78}. Other than the case of Gaussian channels~\cite{Oz84}, currently there is no known single-letter capacity expression for general discrete memoryless MACs (DM-MACs) with feedback. Leveraging the structural results for the optimal encoders for the DSAHT problem, we show that the capacity expression can be thought of as the average per-unit-time reward of an appropriately defined Markov controlled process. In order to achieve this structural result, we introduce some new quantities (other thatn the posterior belief on the message pair from the viewpoint of the receiver that was introduced for the DSAHT problem) that summarize the private beliefs of each transmitter for their own messages conditioned on the corresponding channel input and output.

In the following, we denote random variables with capital letters $X, Y, Z,...$, their realizations with small letters $x, y, z, ...$, and alphabets with calligraphic letters $\cX, \cY, \cZ,...$. A sequence is denoted with $X^1_{1:t} = (X^1_1,...,X^1_t)$.
We use the notation $\P(x|y)$ to denote $\P(X=x|Y=y)$.
The space of probability distributions (or equivalently probability mass functions) on the finite alphabet $\cX$ is denoted by $\cP(\cX)$.

\vspace*{-0.15cm}
\section{Channel Model}
\label{sec:model}

We consider a two-user DM-MAC. The input symbols $X^1$, $X^2$ and the output symbol $Z$ take values in the finite alphabets $\cX^1$, $\cX^2$ and $\cZ$, respectively. The channel is memoryless in the sense that the current channel output is independent of all the past channel inputs and the channel outputs, i.e.,
\begin{equation}
\P(z_t|x^1_{1:t}, x^2_{1:t}, z_{1:t-1}) = Q(z_t|x^1_t, x^2_t).
\end{equation}
Our model considers noiseless feedback, that is, the presence of the channel output $z_{1:t-1}$ to both encoders with unit delay.

Consider the problem of transmission of messages $W^i\in \cW^i=\{1,\ldots,M^i\}, \; i=1,2$, over the MAC with noiseless feedback using fixed length codes of length $n$.
Encoders generate their channel inputs based on their private messages and past outputs. Thus
\begin{align}
X^i_t &= \tilde{f}_t^i(W^i,X^i_{1:t-1},Z_{1:t-1})=f_t^i(W^i,Z_{1:t-1}), \quad i=1,2.
\end{align}
The decoder estimates the messages $W^1$ and $W^2$ based on $n$ channel outputs, $Z_{1:n}$ as
\begin{equation}
(\hat{W}^1,\hat{W}^2) = g(Z_{1:n}).
\end{equation}

A fixed-length transmission scheme for the channel $Q$ is the pair $s=(f,g)$, consisting of the encoding functions
$f=(f^1,f^2)$ with $f^i=f^i_{1:n}$ and decoding function $g$.
The error probability associated with the transmission scheme $s$ is defined as
\begin{equation}
Pe(s) = \P^s((W^1,W^2)\neq (\hat{W}^1,\hat{W}^2)).
\end{equation}

A further generalization of these schemes considers randomized encoding functions, i.e.,
\begin{align}
X^i_t \sim f_t^i(\cdot|W^i,X^i_{1:t-1},Z_{1:t-1}), \qquad i=1,2,
\end{align}
where $f^i_t: \cW^i \times \cX^{t-1}\times \cZ^{t-1}\rightarrow \cP(\cX)$ or even randomized encoding functions with a common randomness (common between the transmitters and the receiver), i.e.,
\begin{align}
X^i_t = f_t^i(W^i,X^i_{1:t-1},Z_{1:t-1},U_t), \qquad i=1,2,
\end{align}
where $\P(u_t|u_{1:t-1},x^1_{1:t-1},x^2_{1:t-1},z_{1:t-1})=\P(u_t)=u(u_t)$, with $u(\cdot)$ the uniform distribution over $[0,1]$. In this case, the decoder is of the form $(\hat{W}^1,\hat{W}^2) = g(Z_{1:n},U_{1:n})$.

For simplicity of exposition we only consider fixed-length schemes, although the model can be generalized to
variable-length schemes and the subsequent structural results are valid in that case as well.

\section{Decentralized sequential active hypothesis testing on the MAC}
\label{sec:dsaht}

One may pose the following optimization problem.
Given the alphabets $\cX^1$, $\cX^2$, $\cZ$, the channel $Q$, the pair $(M_1,M_2)$, and for a fixed length $n$, design the optimal transmission scheme $s=(f,g)$ that minimizes the error probability $P_e(s)$.
\begin{equation}
Pe^* = \min_s Pe(s) \tag{\textbf{P1}}
\end{equation}

In the following we reformulate the problem \textbf{(P1)} into an equivalent optimization problem.
Using the ``common agent'' methodology for decentralized dynamic team problems~\cite{NaMaTe13}, we now decompose the encoding process $X^i_t=f^i_t(W^i,Z_{1:t-1})$ into an equivalent two-stage process.
In the first stage, based on the {common information $Z_{1:t-1}$}, the mappings (or ``partial encoding functions'') $e^i_t$, $i=1,2$ are
generated as $e^i_t=\phi^i_t[z_{1:t-1}]$\footnote{We use square brackets to denote functions with range being function sets, i.e., we use notation $e^i_t=\phi^i_t[z_{1:t-1}]$ because $e^i_t$ is itself a function.} (or collectively, $e_t=(e^1_t,e^2_t)=\phi_t[z_{1:t-1}]$) where $e^i_t : \cW^i \rightarrow \cX^i$. In the second stage, each of these mappings are evaluated at the {private information of each agent}, producing $x^i_t=e^i_t(w^i)$.
In other words, for $i=1,2$, let $\cE^i$ be the collection of  all encoding functions $e^i: \cW^i \rightarrow \mathcal{X}^i$.
In the first stage, the common information given by $Z_{1:t-1}$ is transformed using mappings $\phi^i_t: \cZ^{t-1}  \rightarrow \cE^i$ to produce a pair of encoding functions $e_t=(e^1_t,e^2_t)$. In the second stage these functions are evaluated at the private messages $w^i$ producing $x^i_t=e^i_t(w^i)=\phi^i_t[z_{1:t-1}](w^i)$.

Furthermore, it should be clear that for any pair of encoding functions, the optimal decoder is the ML decoder (assuming equally likely hypotheses), denoted by $g_{ML}$.
Thus we have reformulated problem (\textbf{P1}) as
\begin{equation}
Pe^* = \min_{\phi} Pe(\phi),  \tag{\textbf{P2}}
\end{equation}
where we have defined $Pe(\phi)$ with a slight abuse of notation based on the above equivalence between encoding functions $f$
and mappings $\phi$, as well as the use of ML decoding.

In the following we will show that this problem can be further reformulated as a Markov decision process (MDP).
We define the posterior belief\footnote{Note that the posterior belief is used as a conditional distribution, and as a random variable $\Pi_t(\cdot,\cdot):=\P^{\phi}(W^1=\cdot,W^2=\cdot|Z_{1:t})$} on the message pair at time $t$ as
\begin{subequations}\label{eq:pi}
\begin{align}
\pi_t(w^1,w^2) &\triangleq \P^{f}(W^1=w^1,W^2=w^2|z_{1:t}) \\
 &=\P^{\phi}(W^1=w^1,W^2=w^2|z_{1:t},e_{1:t}).
\end{align}
\end{subequations}
The ML decoder can now be expressed based on $\pi_n$ as
\begin{equation}
(\hat{W}^1,\hat{W}^2) = \arg\max_{w^1,w^2} \Pi_n(w^1,w^2),
\end{equation}
and the resulting error probability is
\begin{equation}
P_e(\phi) =\E^{\phi}[ 1- \max_{w^1,w^2} \Pi_n(w^1,w^2)]=\E^{\phi}[c_{n+1}(\Pi_n)],
\end{equation}
where we defined the terminal cost function as
\begin{equation}\label{eq:terminal}
c_{n+1}(\pi_n) = 1- \max_{w^1,w^2} \pi_n(w^1,w^2),
\end{equation}
and the expectation is wrt the random variable $\Pi_{n}$.

It is now a simple exercise to show that $\pi_t$ can be updated using Bayes rule in a policy-independent way as
\begin{equation}
\pi_t=F(\pi_{t-1},e_t,z_t),
\end{equation}
where the mapping $F$ is defined through
\begin{subequations}\label{eq:pi_update}
\begin{align}
\pi_t&(w^1,w^2) \nonumber \\
&= \P^{\phi}(w^1,w^2|z_{1:t},e_{1:t}) \\
 &=\frac{\P^{\phi}(w^1,w^2,z_t,e_t|z_{1:t-1},e_{1:t-1})}{\P^{\phi}(z_t,e_t|z_{1:t-1},e_{1:t-1})} \\
 &=\frac{\P^{\phi}(z_t|w^1,w^2,z_{1:t-1},e_{1:t})\pi_{t-1}(w^1,w^2)}{\P^{\phi}(z_t|z_{1:t-1},e_{1:t})} \\
 &=\frac{Q(z_t|e^1_t(w^1),e^2_t(w^2))\pi_{t-1}(w^1,w^2)}
        {\sum\limits_{\tw^1,\tw^2} Q(z_t|e^1_t(\tw^1),e^2_t(\tw^2))\pi_{t-1}(\tw^1,\tw^2)}.
\end{align}
\end{subequations}
We summarize the above result into the following lemma.
\begin{lemma}
The posterior belief $\pi_t$ on the message pair $(W^1,W^2)$ can be updated in a policy-independent (i.e., $\phi$-independent) way as $\pi_t=F(\pi_{t-1},e_t,z_t)$.
\end{lemma}
\begin{IEEEproof}
The proof is essentially given in~\eqref{eq:pi_update}.
\end{IEEEproof}

The final step in the ``common agent'' methodology is to show that a fictitious common agent who observes only the common information $Z_{1:t-1}$ faces an MDP with state at time $t$, $\Pi_{t-1}$; action $E_t=(E^1_t,E^2_t)$; zero instantaneous costs $c_t(\Pi_{t-1},E_t)=0$ for $t=1,\ldots,n$; and terminal cost $c_{n+1}(\Pi_n)$.
Indeed, $(\Pi_{t-1},E_t)_{t\geq 1}$ is a controlled Markov chain, since
\begin{subequations}
\begin{align}
\P^{\phi}&(\pi_t|\pi_{1:t-1},e_{1:t}) \nonumber \\
 &= \sum_{z_t} \P^{\phi}(\pi_t|z_t, \pi_{1:t-1},e_{1:t}) \times \nonumber \\
  &\qquad \sum_{w^1,w^2} \P^{\phi}(z_t|w^1,w^2, \pi_{1:t-1},e_{1:t}) \times \nonumber \\
   &\qquad\qquad\qquad\P^{\phi}(w^1,w^2| \pi_{1:t-1},e_{1:t})\\
 &= \sum_{z_t} 1_{F(\pi_{t-1},e_t,z_t)}(\pi_t) \times \nonumber \\
  &\qquad \sum_{w^1,w^2} Q(z_t|e_t^1(w^1),e_t^2(w^2)) \pi_{t-1}(w^1,w^2) \\
 &= \P(\pi_t|\pi_{t-1},e_{t}).
\end{align}
\end{subequations}

At this point we have transformed problem (\textbf{P2}) into the following MDP
\begin{equation}
Pe^* = \min_{\phi} \E [ \sum_{t=1}^n c_t(\Pi_{t-1},E_t) +  c_{n+1}(\Pi_n) ]. \tag{\textbf{P3}}
\end{equation}

As a result, the optimal policy is deterministic Markovian, i.e., of the form $E_t=\theta_t[\Pi_{t-1}]$ (or explicitly, $E^i_t=\theta^i_t[\Pi_{t-1}]$), resulting in an encoding policy of the form
$X^i_t=\theta^i_t[\Pi_{t-1}](W^i)=f^i_t(\Pi_{t-1},W^i)$.

Furthermore, the characterization of the optimal Markov policy is the backward dynamic program
\begin{subequations}\label{eq:bdp}
\begin{align}
V_{n+1}(\pi_n) &= c_{n+1}(\pi_n) \\
V_t(\pi_{t-1}) &= \min_{e_t}\E [  V_{t+1}(F(\pi_{t-1},e_t,Z_t)) | \pi_{t-1} , e_t ] \\
               &= \min_{e_t} \sum_{z_t,w^1,w^2} Q(z_t|e_t^1(w^1),e_t^2(w^2)) \pi_{t-1}(w^1,w^2) \nonumber \\
               &\qquad\qquad \qquad\qquad V_{t+1}(F(\pi_{t-1},e_t,z_t)).
\end{align}
\end{subequations}

All the above results can be summarized in the following theorem
\begin{theorem}\label{th:active}
The optimization problem~\textbf{(P1)} can be restated as an MDP with state at time $t$, $\Pi_{t-1}$; action $E_t=(E^1_t,E^2_t)$; zero instantaneous costs $c_t(\Pi_{t-1},E_t)=0$ for $t=1,\ldots,n$; and terminal cost $c_{n+1}(\Pi_n)$ given in~\eqref{eq:terminal}.
Consequently, the optimal encoders are of the form $X^i_t=E^i_t(W^i)=\theta^i_t[\Pi_{t-1}](W^i)=f^i_t(\Pi_{t-1},W^i)$. Finally, the mapping
$\theta$ can be found through backward dynamic programming as in~\eqref{eq:bdp}.
\end{theorem}
\begin{IEEEproof}
The proof is given in the previous discussion.
\end{IEEEproof}

We conclude this section by pointing out that the main idea behind the characterization of the optimal solution of the decentralized sequential active hypothesis testing (DSAHT) problem was to transform the decentralized problem (three agents with common and private information) into a centralized problem (single, ``fictitious'' agent) who observes the common information, $Z_{1:t-1}$ of all three agents and takes actions $E_t=(E^1_t,E^2_t)$ which are then evaluated on the private information $W^i$ to generate the inputs $X^i_t$. The price to pay for this reduction is that the action set of the fictitious common agent is now a pair of functions (instead of the transmitted symbols). The gain from this characterization is that the solution can be obtained by backward dynamic programming and the resulting optimal encoding functions do not have a time-varying domain, but can be summarized into a sufficient statistic $\Pi_{t-1}$.

\optv{arxiv}{
\subsection{Alternative Objectives and formulations}
}

\optv{2col}{
The same structural results can be derived for similar problems where the terminal cost is not the one defined above but an arbitrary function of $\pi_n$. Due to space limitations, these alternative formulations are not presented here. They can be found in the full version of the paper~\cite{AnPr20arxiv}.
}

\optv{arxiv}{

The same structural results can be derived for similar problems where the terminal cost is not the one defined above but an arbitrary function of $\pi_n$. We mention here three such interesting cases
\begin{enumerate}
\item The first one relates to the entropy $H(W^1,W^2|Z_{1:n})$ or equivalently the negative of the mutual information $I(W^1,W^2;Z_{1:n})$.
\begin{subequations}
\begin{align}
\E [-\log \Pi_n(W^1,W^2)] &= \E[ -\sum_{w^1,w^2} \Pi_n(w^1,w^2) \log \Pi_n(w^1,w^2)  ] \\
 &= H(W^1,W^2|Z_{1:n}) \\
 &= H(W^1,W^2) - I(W^1,W^2;Z_{1:n})
\end{align}
\end{subequations}

\item The second one relates to the conditional entropy $H(W^1|W^2,Z_{1:n})$ or equivalently the negative of the mutual information $I(W^1;Z_{1:n}|W^2)$.
\begin{subequations}
\begin{align}
\E [-\log \Pi_n(W^1|W^2)] &= \E[ -\sum_{w^1,w^2} \Pi_n(w^1,w^2) \log \Pi_n(w^1|w^2)  ] \\
 &= H(W^1|W^2,Z_{1:n}) \\
 &= H(W^1|W^2) - I(W^1;Z_{1:n}|W^2).
\end{align}
\end{subequations}

\item The last one relates to the log-likelihood ratio of the true message pair
\begin{equation}
\E [-\log \frac{\Pi_n(W^1,W^2)}{1-\Pi_n(W^1,W^2)}] = \E[ -\sum_{w^1,w^2} \Pi_n(w^1,w^2) \log \frac{\Pi_n(w^1,w^2)}{1-\Pi_n(w^1,w^2)}  ].
\end{equation}
\end{enumerate}
Interestingly, in the above cases the problem can be reformulated so that the terminal cost is distributed into time-invariant instantaneous costs throughout the transmission, with these instantaneous costs having an intuitive explanation. Indeed,
we can define time-invariant instantaneous cost functions $c(\pi_{t-1},e_t)$ for $t=1,\ldots,n$ and eliminate the terminal cost $c_{n+1}(\pi_n)$ as follows
\begin{enumerate}
\item
\begin{subequations}
\begin{align}
\E [-\log \Pi_n(W^1,W^2)] &= \E[-\log\Pi_{0}(W^1,W^2)] + \sum_{t=1}^n \E[-\log \frac{\Pi_t(W^1,W^2)}{\Pi_{t-1}(W^1,W^2)}]\\
 &= H(W^1,W^2) + \sum_{t=1}^n -\E[\log \frac{\Pi_t(W^1,W^2)}{\Pi_{t-1}(W^1,W^2)}]
\end{align}
where
\begin{align}
-\E&[\log \frac{\Pi_t(W^1,W^2)}{\Pi_{t-1}(W^1,W^2)}] \\
 &= -\E[\log \frac{F(\Pi_{t-1},E_t,Z_t)(W^1,W^2)}{\Pi_{t-1}(W^1,W^2)}] \\
 &= -\E[\log\frac{Q(Z_t|E^1_t(W^1),E^2_t(W^2))}{\sum_{\tw^1,\tw^2} Q(Z_t|E^1_t(\tw^1),E^2_t(\tw^2))\Pi_{t-1}(\tw^1,\tw^2)}] \\
 &= -\E[\sum_{z_t,w^1,w^2}Q(z_t|E^1_t(w^1),E^2_t(w^2))\Pi_{t-1}(w^1,w^2) \nonumber \\
   & \qquad \qquad\log\frac{Q(z_t|E^1_t(w^1),E^2_t(w^2))}{\sum_{\tw^1,\tw^2} Q(z_t|E^1_t(\tw^1),E^2_t(\tw^2))\Pi_{t-1}(\tw^1,\tw^2)}] \\
 &= -I(W^1,W^2;Z_t|Z_{1:t-1}) \\
 &= \E[c(\Pi_{t-1},E_t)],
\end{align}
with
\begin{align}
c(\pi,e) &= -\sum_{z,w^1,w^2}Q(z|e^1(w^1),e^2(w^2))\pi(w^1,w^2) \log\frac{Q(z|e^1(w^1),e^2(w^2))}{\sum_{\tw^1,\tw^2} Q(z|e^1(\tw^1),e^2(\tw^2))\pi(\tw^1,\tw^2)}].
\end{align}
\end{subequations}
As a result, minimizing the final entropy $H(W^1,W^2|Z_{1:n})$ (or equivalently, maximizing the final mutual information $I(W^1,W^2;Z_{1:n})$) is equivalent to minimizing the cumulative conditional entropy $I(W^1,W^2;Z_t|Z_{1:t-1})$ which is also equivalent to maximizing (on the average) of the cumulative drift of the log-likelihood of the true message pair
$\log \Pi_t(W^1,W^2)$.

\item
\begin{subequations}
\begin{align}
\E [-\log \Pi_n(W^1|W^2)] &= \E[-\log\Pi_{0}(W^1|W^2)] + \sum_{t=1}^n \E[-\log \frac{\Pi_t(W^1|W^2)}{\Pi_{t-1}(W^1|W^2)}]\\
 &= H(W^1|W^2) + \sum_{t=1}^n -\E[\log \frac{\Pi_t(W^1|W^2)}{\Pi_{t-1}(W^1|W^2)}]
\end{align}
where
\begin{align}
-\E&[\log \frac{\Pi_t(W^1|W^2)}{\Pi_{t-1}(W^1|W^2)}] \\
 &= -\E[\log \frac{F(\Pi_{t-1},E_t,Z_t)(W^1|W^2)}{\Pi_{t-1}(W^1|W^2)}] \\
 &= -\E[\log\frac{Q(Z_t|E^1_t(W^1),E^2_t(W^2))}
 {\sum_{\tw^1} Q(Z_t|E^1_t(\tw^1),E^2_t(W^2))\Pi_{t-1}(\tw^1|W^2)}] \\
 &= -\E[\sum_{z_t,w^1,w^2}Q(z_t|E^1_t(w^1),E^2_t(w^2))\Pi_{t-1}(w^1,w^2) \nonumber \\
       & \qquad \qquad \log\frac{Q(z_t|E^1_t(w^1),E^2_t(w^2))}
   {\sum_{\tw^1} Q(z_t|E^1_t(\tw^1),E^2_t(w^2))\Pi_{t-1}(\tw^1|w^2)}] \\
 &= -I(W^1;Z_t|W^2,Z_{1:t-1}) \\
 &= \E[c(\Pi_{t-1},E_t)],
\end{align}
with
\begin{align}
c(\pi,e) &= -\sum_{z,w^1,w^2}Q(z|e^1(w^1),e^2(w^2))\pi(w^1,w^2) \log\frac{Q(z|e^1(w^1),e^2(w^2))}
 {\sum_{\tw^1} Q(z|e^1(\tw^1),e^2(w^2))\pi(\tw^1|w^2)}].
\end{align}
\end{subequations}

\item
\begin{subequations}
\begin{align}
\E &[-\log \frac{\Pi_n(W^1,W^2)}{1-\Pi_n(W^1,W^2)}] \\
 &= \E [-\log \frac{\Pi_0(W^1,W^2)}{1-\Pi_0(W^1,W^2)}] + \sum_{t=1}^n  \E [-\log \frac{\Pi_t(W^1,W^2)(1-\Pi_{t-1}(W^1,W^2))}{(1-\Pi_{t}(W^1,W^2))\Pi_t(W^1,W^2)}]
 \end{align}
 where we identify the terms inside the summation as
\begin{align}
 \E& [-\log \frac{\Pi_t(W^1,W^2)(1-\Pi_{t-1}(W^1,W^2))}{(1-\Pi_{t}(W^1,W^2))\Pi_t(W^1,W^2)}] \\
 &= \E[-\log\frac{Q(Z_t|E^1_t(W^1),E^2_t(W^2))}{\sum\limits_{(\tw^1,\tw^2)\neq (W^1,W^2)}\frac{\Pi_{t-1}(\tw^1,\tw^2)}{1-\Pi_{t-1}(W^1,W^2)}Q(Z_t|E^1_t(W^1),E^2_t(W^2))}] \\
 &= \E[-\sum_{z_t,w^1,w^2}Q(z_t|E^1_t(w^1),E^2_t(w^2))\Pi_{t-1}(w^1,w^2) \nonumber \\
  &\qquad\qquad\qquad \log\frac{Q(z_t|E^1_t(w^1),E^2_t(w^2))}{\sum\limits_{(\tw^1,\tw^2)\neq (w^1,w^2)}\frac{\Pi_{t-1}(\tw^1,\tw^2)}{1-\Pi_{t-1}(w^1,w^2)}Q(z_t|E^1_t(w^1),E^2_t(w^2))}] \\
 &=\E[c(\Pi_{t-1},E_t)]
  \end{align}
 with
  \begin{align}
 c&(\pi,e) \nonumber \\
 &=-\sum_{z,w^1,w^2}Q(z|e^1(w^1),e^2(w^2))\pi(w^1,w^2) \nonumber \\
  &\qquad\qquad\qquad \log\frac{Q(z|e^1(w^1),e^2(w^2))}{\sum\limits_{(\tw^1,\tw^2)\neq (w^1,w^2)}\frac{\pi(\tw^1,\tw^2)}{1-\pi(w^1,w^2)}Q(z|e^1(w^1),e^2(w^2))} \\
  &=-\sum_{w^1,w^2}\pi(w^1,w^2) D(Q(\cdot|e^1(w^1),e^2(w^2))||\sum_{(\tw^1,\tw^2)\neq (w^1,w^2)}\frac{\pi(\tw^1,\tw^2)}{1-\pi(w^1,w^2)}Q(\cdot|e^1(w^1),e^2(w^2)) \\
  &=-EJS(\pi,\{Q(\cdot|e^1(w^1),e^2(w^2))\}_{(w^1,w^2)\in\cW^1\times\cW^2}),
\end{align}
\end{subequations}
where $EJS$ denotes the extrinsic Jensen-Shannon divergence~\cite{NaJaWi15}.
\end{enumerate}

Clearly one may consider other cost functions, e.g., a linear combination of 1) and 2) or even a linear combination of 1), 2), and the symmetric quantity $H(W^2|W^1,Z_{1:n})$. Similarly, one can consider linear combination of log-likelihood ratios such as the one appearing in 3) with conditional beliefs $\Pi_n(W^1|W^2)$, or $\Pi_n(W^2|W^1)$ in place of the joint belief $\Pi_n(W^1,W^2)$, resulting in time-invariant instantaneous costs with appropriate EJS-related quantities.

Since the reformulated problem involves time-invariant costs and a time-homogenous controlled Markov process, we can extend these results to infinite-horizon formulations with either discounted reward or average reward per unit time. The optimal policy will also be time-invariant in this case and it is characterized through the solution of the following fixed-point equations. For instance,
for the average reward per unit time we have
\begin{subequations}
\begin{align}
J+V(\pi)  &= \min_{e} c(\pi,e) +  \E[ V(F(\pi,e,Z)) | \pi , e ] \\
          &= \min_{e} c(\pi,e) +  \sum_{z,w^1,w^2} Q(z|e^1(w^1),e^2(w^2))\pi(w^1,w^2) V(F(\pi,e,z)).
\end{align}
\end{subequations}

We remark at this point, that a similar formulation with infinite horizon and variable length coding where we minimize a linear combination of the error probability and the length of transmission results in exactly the same structural results, i.e., summarizing the common history $Z_{1:t}$ into the belief $\Pi_{t}$ and in addition has time-invariant optimal solutions.
This formulation is the decentralized equivalent of the point-to-point active sequential hypothesis testing discussed in~\cite{NaJa13}.

} 

\section{Connection between DSAHT and the MAC channel capacity}

\subsection{Multi-letter capacity expressions}

A multi-letter capacity expression for DM-MAC with noiseless feedback has been established in~\cite{Kr98} and can be stated as follows.

\begin{fact}[Theorem 5.1 in \cite{Kr98}, \cite{Sa78}]\label{capexp}
The capacity region of the DM-MAC with feedback is
$\cC_{FB}  =\bigcup_{n=1}^{\infty}\cC_{n}$
where $\cC_{n}$, the {directed information $n$-th inner bound region}, is defined as $\cC_n = \text{co}\left(\cR_n\right)$, where $co(A)$ denotes the convex hull of a set $A$, and
\vspace*{-0.25cm}
\begin{align}
\cR_n = \cup_{\cP_n}\{ (R_1,&R_2):
 0 \leq R_1 \leq I_{n}(X^1\rightarrow Z||X^2),  \nonumber \\
& 0 \leq R_2 \leq I_{n}(X^2\rightarrow Z||X^1),\nonumber \\
& 0 \leq R_1+R_2 \leq I_{n}(X^1,X^2\rightarrow Z)
\},
\end{align}
where $I_n(A\rightarrow B||C) = \frac{1}{n}\sum_{t=1}^n I(A_{1:t};B_t|C_{1:t},B_{1:t-1})= \frac{1}{n}\sum_{t=1}^n I(A_{t};B_t|C_{1:t},B_{1:t-1})$.
All information quantities are evaluated using the joint distribution
\begin{align}
\P(x^1_{1:n},x^2_{1:n},z_{1:n}) = \prod_{t=1}^n Q(z_t|&x^1_t,x^2_t)q^1_t(x^1_t|x^1_{1:t-1},z_{1:t-1})\times \nonumber \\
& q^2_t(x^2_t|x^2_{1:t-1},z_{1:t-1}),
\end{align}
and the union is over all input joint distributions on $x^1_t, x^2_t$ that are conditionally factorizable as
\begin{align}
\label{eq:optdist}
&\P(x^1_t,x^2_t|x^1_{1:t-1},x^2_{1:t-1},z_{1:t-1}) = \nonumber \\
 & \quad q^1_t(x^1_t|x^1_{1:t-1},z_{1:t-1})
 q^2_t(x^2_t|x^2_{1:t-1},z_{1:t-1})
\end{align}
for $t = 1,2,...,n$.

Furthermore, the regions $\cC_{n}$ can be expressed in the form~\cite{Sa78}
\begin{align}
\label{eq:Tinnerbound}
\cC_n &= \left\{(R_1,R_2)\geq0:\forall\ \underline{\lambda}=(\lambda_1,\lambda_2,\lambda_3)\in \Real^3_+, \right.  \nonumber \\
& \qquad \left. \lambda_1R_1+\lambda_2R_2+\lambda_3(R_1+R_2)\leq C_n(\underline{\lambda})\right\},
\end{align}
where
\begin{subequations}
\label{eq:region_to_cost_fb}
\begin{align}
C_n(\underline{\lambda}) &\triangleq \sup_{\cP_n} I_n(\underline{\lambda})\\
I_n(\underline{\lambda}) &\triangleq
\lambda_1 I_{n}(X^1\rightarrow Z||X^2)+
\lambda_2 I_{n}(X^2\rightarrow Z||X^1)+ \nonumber \\
&\qquad \lambda_3 I_{n}(X^1,X^2\rightarrow Z) \\
 & =\frac{1}{n} \sum_{t=1}^n
 [  \lambda_1 I(X^1_t;Z_t|X^2_{1:t},Z_{1:t-1}) + \nonumber \\
 & \qquad\qquad \lambda_2 I(X^2_t;Z_t|X^1_{1:t},Z_{1:t-1}) + \nonumber \\
 & \qquad\qquad  \lambda_3 I(X^1_t,X^2_t;Z_t|Z_{1:t-1})]
\end{align}
\end{subequations}
and in the above, the set $\cP_n$ is defined as
\begin{align}
\cP_n = \left\{ (q^1_t,q^2_t)_{t=1,\ldots,n}: q^i_t \in  (\cX^i)^{t-1} \times \cZ^{t-1} \rightarrow \cP(\cX^i)   \right\}.
\end{align}
\end{fact}

Observe that the problem of evaluating capacity is essentially (at least) as hard as the problem of evaluating the quantity $C_n(\underline{\lambda})$ for a given $\underline{\lambda}$. Also note that
the optimization problem involved in evaluating $C_n(\underline{\lambda})$ can be thought of as a decentralized optimization problem involving two agents: the first is choosing the distribution $q^1_t$ on $x^1_t$
after observing the common information $z_{1:t-1}$ and his private information $x^1_{1:t-1}$, while the second
is choosing the distribution $q^2_t$ on $x^2_t$
after observing the common information $z_{1:t-1}$ and his private information $x^2_{1:t-1}$. This decentralized nature contributes to the difficulty of this optimization problem.

\subsection{Input distributions induced by structured strategies}

How can the DSAHT problem stated in the previous section, together with the structural results obtained, help us with the problem of characterizing the feedback capacity for the DM-MAC?
The idea behind the answer is that in evaluating the capacity of the DM-MAC one may restrict attention to the optimal encoders obtained for the DSAHT problem without loss of optimality. Indeed, Theorem~\ref{th:active}
states that transmitters of the form $X^i_t=E^i_t(W^i)=\theta^i_t[\Pi_{t-1}](W^i)=f^i_t(\Pi_{t-1},W^i)$ are sufficient for minimizing the error probability of the message pair. In the following we show that the information theoretic quantities involved in the evaluation of $I_n(\underline{\lambda})$ in~\eqref{eq:region_to_cost_fb}, as well as the input distributions $q^i_t$ in~\eqref{eq:optdist} take a specific simplified form when the structured strategies of Theorem~\ref{th:active} are used.

To aid this goal,
we define additional posterior beliefs on the message at time $t$ given all available information to each transmitter as
\begin{equation}
\hat{\pi}^i_t(w^i) \triangleq \P(W^i=w^i|x^i_{1:t},z_{1:t}), \qquad i=1,2.
\end{equation}
Note that $\hat{\pi}^i_t$ is the marginal belief that user $i$ maintains on her own message $W^i$.
We now state the following lemma regarding the induced distributions $\P^{\theta}(x^i_t|x^i_{1:t-1},z_{1:t-1})$.
\begin{lemma}\label{lm:distributions}
The conditional distribution $\P^{\theta}(x^1_t,x^2_t|x^1_{1:t-1},x^2_{1:t-1},z_{1:t-1})$,
induced by structured strategies of the form $X^i_t=E^i_t(W^i)=\theta^i_t[\Pi_{t-1}](W^i)=f^i_t(\Pi_{t-1},W^i)$ is always in $\cP_n$, i.e., it can be factored as
\begin{align}
\P^{\theta}&(x^1_t,x^2_t|x^1_{1:t-1},x^2_{1:t-1},z_{1:t-1}) \nonumber \\
&= \P^{\theta}(x^1_t|x^1_{1:t-1},z_{1:t-1})\P^{\theta}(x^2_t|x^2_{1:t-1},z_{1:t-1}).
\end{align}
Furthermore, the marginal distributions $\P^{\theta}(x^i_t|x^i_{1:t-1},z_{1:t-1})$ can be simplified as
\begin{align}\label{eq:simp2}
&\P^{\theta}(x^i_t|x^i_{1:t-1},z_{1:t-1})
 = \sum_{w^i} 1_{e^i_t(w^i)}(x^i_t) \hat{\pi}^i_{t-1}(w^i).
\end{align}
\end{lemma}
\begin{IEEEproof}
We first show (using induction) that the conditional distribution $\P^{\theta}(w^1,w^2|x^1_{1:t},x^2_{1:t},z_{1:t})$ induced by structured strategies of the form $X^i_t=E^i_t(W^i)=\theta^i_t[\Pi_{t-1}](W^i)=f^i_t(\Pi_{t-1},W^i)$ can be factored as
\begin{subequations}
\begin{align}
\P^{\theta}(w^1,w^2|x^1_{1:t},x^2_{1:t},z_{1:t})
 &=\P^{\theta}(w^1|x^1_{1:t},z_{1:t})\P^{\theta}(w^2|x^2_{1:t},z_{1:t}) \\
 &=\hat{\pi}^1_t(w^1)\hat{\pi}^2_t(w^2).
\end{align}
\end{subequations}
Indeed, for $t=0$ we have $\P^{\theta}(w^1,w^2)=(1/M^1)(1/M^2)=\P(w^1)\P(w^2)=\hat{\pi}^1_0(w^1)\hat{\pi}^2_0(w^2)$.
Assuming that the above is true for time $t-1$ we have for $t$
\begin{subequations}
\begin{align}
&\P^{\theta}(w^1,w^2|x^1_{1:t},x^2_{1:t},z_{1:t}) \nonumber \\
 =&\frac{\P^{\theta}(w^1,w^2,x^1_t,x^2_t,z_t|x^1_{1:t-1},x^2_{1:t-1},z_{1:t-1})}
        {\sum\limits_{\tw^1,\tw^2}\P^{\theta}(\tw^1,\tw^2,x^1_t,x^2_t,z_t|x^1_{1:t-1},x^2_{1:t-1},z_{1:t-1})} \\
 =&\frac{Q(z_t|x^1_t,x^2_t)1_{e^1_t(w^1)}(x^1_t)1_{e^2_t(w^2)}(x^2_t)\hat{\pi}^1_{t-1}(w^1)\hat{\pi}^2_{t-1}(w^2)}
        {\sum\limits_{\tw^1,\tw^2}Q(z_t|x^1_t,x^2_t)1_{e^1_t(\tw^1)}(x^1_t)1_{e^2_t(\tw^2)}(x^2_t)\hat{\pi}^1_{t-1}(\tw^1)\hat{\pi}^2_{t-1}(\tw^2)} \\
 =&\frac{1_{e^1_t(w^1)}(x^1_t)1_{e^2_t(w^2)}(x^2_t)\hat{\pi}^1_{t-1}(w^1)\hat{\pi}^2_{t-1}(w^2)}
        {\sum\limits_{\tw^1,\tw^2}1_{e^1_t(\tw^1)}(x^1_t)1_{e^2_t(\tw^2)}(x^2_t)\hat{\pi}^1_{t-1}(\tw^1)\hat{\pi}^2_{t-1}(\tw^2)} \\
 =&\frac{1_{e^1_t(w^1)}(x^1_t)\hat{\pi}^1_{t-1}(w^1)}
        {\sum\limits_{\tw^1}1_{e^1_t(\tw^1)}(x^1_t)\hat{\pi}^1_{t-1}(\tw^1)}
   \frac{1_{e^2_t(w^2)}(x^2_t)\hat{\pi}^2_{t-1}(w^2)}
        {\sum\limits_{\tw^2}1_{e^2_t(\tw^2)}(x^2_t)\hat{\pi}^2_{t-1}(\tw^2)} \\
 =&\P^{\theta}(w^1|x^1_{1:t},z_{1:t})\P^{\theta}(w^2|x^2_{1:t},z_{1:t})  \\
 =&\hat{\pi}^1_t(w^1)\hat{\pi}^2_t(w^2).
\end{align}
\end{subequations}
As a byproduct of this proof we see that the belief $\hat{\pi}^i_t$ can be updated as
\begin{align}
\hat{\pi}^i_t(w^i) = \frac{1_{e^i_t(w^i)}(x^i_t)\hat{\pi}^i_{t-1}(w^i)}{\sum_{\tw^i} 1_{e^i_t(\tw^i)}(x^i_t)\hat{\pi}^i_{t-1}(\tw^i)},
\end{align}
or more succinctly
\begin{align}
\hat{\pi}^i_t &= \hat{F}^i(\hat{\pi}^i_{t-1},e^i_t,x^i_t)= \hat{F}^i(\hat{\pi}^i_{t-1},\theta^i_t[\pi_{t-1}],x^i_t), \quad i=1,2 .
\end{align}

Now the induced distributions $\P^{\theta}(x^1_t,x^2_t|x^1_{1:t-1},x^2_{1:t-1},z_{1:t-1})$ can be evaluated as
\begin{subequations}
\begin{align}
&\P^{\theta}(x^1_t,x^2_t|x^1_{1:t-1},x^2_{1:t-1},z_{1:t-1}) \nonumber \\
 &= \sum_{w^1,w^2} \P^{\theta}(x^1_t,x^2_t,w^1,w^2|x^1_{1:t-1},x^2_{1:t-1},z_{1:t-1}) \\
 &= \sum_{w^1,w^2} \P^{\theta}(x^1_t,x^2_t|w^1,w^2,x^1_{1:t-1},x^2_{1:t-1},z_{1:t-1}) \nonumber \\
 &\qquad\qquad \P^{\theta}(w^1,w^2|x^1_{1:t-1},x^2_{1:t-1},z_{1:t-1}) \\
 &= [\sum_{w^1} 1_{e^1_t(w^1)}(x^1_t)\hat{\pi}^1_{t-1}(w^1)][\sum_{w^2} 1_{e^2_t(w^2)}(x^2_t)\hat{\pi}^2_{t-1}(w^2)] \\
 &= \P(x^1_t|\hat{\pi}^1_{t-1},e^1_t)\P(x^2_t|\hat{\pi}^2_{t-1},e^2_t)\\
 &= \P(x^1_t|\hat{\pi}^1_{t-1},\theta^1_t[\pi_{t-1}]) \P(x^2_t|\hat{\pi}^2_{t-1},\theta^2_t[\pi_{t-1}])\\
 &= \P^{\theta^1_t}(x^1_t|\hat{\pi}^1_{t-1},\pi_{t-1}) \P^{\theta^2_t}(x^2_t|\hat{\pi}^2_{t-1},\pi_{t-1}).
\end{align}
\end{subequations}
The last equation is the proof that $\P^{\theta}(x^1_t,x^2_t|x^1_{1:t-1},x^2_{1:t-1},z_{1:t-1})$ factors into
the conditionals $\P^{\theta}(x^i_t|x^i_{1:t-1},z_{1:t-1})$ and that in the latter expressions the conditional history
$x^i_{1:t-1},z_{1:t-1}$ is summarized in the quantities $(\hat{\pi}^i_{t-1},e^i_t)=(\hat{\pi}^i_{t-1},\theta^i_t[\pi_{t-1}])=(\hat{\pi}^i_{t-1},\pi_{t-1})$.
\end{IEEEproof}

The next step in the development is to derive simplified expressions for the mutual information quantities that are involved in the $I_n(\underline{\lambda})$ in ~\eqref{eq:region_to_cost_fb}. Specifically, we will derive simplified expressions for the quantities $I(X^1_t;Z_t|X^2_{1:t},Z_{1:t-1})$, $I(X^2_t;Z_t|X^1_{1:t},Z_{1:t-1})$, and $I(X^1_t,X^2_t;Z_t|Z_{1:t-1})$, or equivalently, for the quantities $H(Z_t|X^2_{1:t},Z_{1:t-1})$, $H(Z_t|X^1_{1:t},Z_{1:t-1})$,
$H(Z_t|Z_{1:t-1})$ and $H(Z_t|X^1_t,X^2_t)$. Our results are summarized in the following theorem.
\begin{theorem}\label{th:capacity}
The mutual information quantities that are involved in the expression for $I_n(\underline{\lambda})$ in ~\eqref{eq:region_to_cost_fb} can be evaluated as expectations of time invariant quantities depended only on $\Pi_{t-1}$, $\hat{\Pi}^i_{t-1}$ and $E_t$. Specifically, for each $t=1,\ldots,n$ we have
\begin{subequations}
\begin{align}
\label{eq:th2a}
I(X^1_t;Z_t|X^2_{1:t},Z_{1:t-1}) &= \E^{\theta}[ i_1(\hat{\Pi}^2_{t-1},\Pi_{t-1},E_t) ] \\
I(X^2_t;Z_t|X^1_{1:t},Z_{1:t-1}) &= \E^{\theta}[ i_2(\hat{\Pi}^1_{t-1},\Pi_{t-1},E_t) ] \\
I(X^1_t,X^2_t;Z_t|Z_{1:t-1}) &= \E^{\theta}[ i_3(\Pi_{t-1},E_t) ],
\end{align}
\end{subequations}
where the functions $i_1$, $i_2$, $i_3$ are specified in the proof of the theorem and
expectations are taken wrt the joint distribution
\begin{subequations}
\begin{align}
&\P^{\theta}(\pi_{0:n-1},\hat{\pi}_{0:n-1},e_{1:n}) \nonumber \\
&= \prod_{t=0}^{n-1} \P^{\theta}(\pi_t,\hat{\pi}_t,e_{t+1}|\pi_{0:t-1},\hat{\pi}_{0:t-1},e_{1:t}) \\
&=
\prod_{t=0}^{n-1} 1_{\theta_{t+1}[\pi_t]}(e_{t+1}) \sum_{z_t,x^1_t,x^2_t}
  Q(z_t|x^1_t,x^2_t) 1_{F(\pi_{t-1},e_t,z_t)}(\pi_t) \nonumber \\
  &\quad  1_{\hat{F}^1(\hat{\pi}^1_{t-1},e^1_t,x^1_t)}(\hat{\pi}^1_t) 1_{\hat{F}^2(\hat{\pi}^2_{t-1},e^2_t,x^2_t)}(\hat{\pi}^2_t) \nonumber \\
  &\quad \sum_{w^1,w^2}1_{e^1_t(w^1)}(x^1_t)1_{e^2_t(w^2)}(x^2_t) \hat{\pi}^1_{t-1}(w^1)\hat{\pi}^2_{t-1}(w^2).
\end{align}
\end{subequations}
\end{theorem}

\begin{IEEEproof}
\optv{2col}{
The proof is omitted due to space limitations. It can be found in the full version of the paper in~\cite{AnPr20arxiv}.
}
\optv{arxiv}{

Let us first consider the quantity $I(X^1_t;Z_t|X^2_{1:t},Z_{1:t-1})=H(Z_t|X^2_{1:t},Z_{1:t-1})-H(Z_t|X^1_t,X^2_t)$.
For the quantity $H(Z_t|X^2_{1:t},Z_{1:t-1})$ we have
\begin{subequations}
\begin{align}\label{eq:h1_a}
&H(Z_t|X^2_{1:t},Z_{1:t-1}) \nonumber \\
 &= - \sum_{x^2_{1:t-1},z_{1:t-1}}\P(x^2_{1:t-1},z_{1:t-1})\sum_{x^2_t}\P(x^2_t|x^2_{1:t-1},z_{1:t-1})  \nonumber \\
  &\quad \sum_{z_t}\P(z_t|x^2_{1:t},z_{1:t-1}) \log \P(z_t|x^2_{1:t},z_{1:t-1}),
\end{align}
where $\P(x^2_t|x^2_{1:t-1},z_{1:t-1})$ is given by~\eqref{eq:simp2} in Lemma~\ref{lm:distributions}
and
\begin{align}
&\P(z_t|x^2_{1:t},z_{1:t-1}) \nonumber \\
 &= \sum_{x^1_t} Q(z_t|x^1_t,x^2_t) \P(x^1_t|x^2_{1:t},z_{1:t-1}) \label{eq:h1_b}\\
 &= \sum_{x^1_t} Q(z_t|x^1_t,x^2_t) \sum_{w^1,w^2} \P(x^1_t,w^1,w^2|x^2_{1:t},z_{1:t-1}) \\
 &= \sum_{x^1_t} Q(z_t|x^1_t,x^2_t) \sum_{w^1,w^2} 1_{e^1_t(w^1)}(x^1_t)\P(w^1,w^2|x^2_{1:t},z_{1:t-1}) \\
 &= \sum_{x^1_t} Q(z_t|x^1_t,x^2_t) \sum_{w^1,w^2} 1_{e^1_t(w^1)}(x^1_t) \nonumber \\
  &\qquad\qquad \frac{\P(w^1,w^2,x^2_t|x^2_{1:t-1},z_{1:t-1})}{\sum_{w^1,w^2}\P(w^1,w^2,x^2_t|x^2_{1:t-1},z_{1:t-1})} \\
 &= \sum_{x^1_t} Q(z_t|x^1_t,x^2_t) \sum_{w^1,w^2} 1_{e^1_t(w^1)}(x^1_t) \nonumber \\
  &\frac{1_{e^2_t(w^2)}(x^2_t)\P(w^1|w^2,x^2_{1:t-1},z_{1:t-1})\P(w^2|x^2_{1:t-1},z_{1:t-1})}
        {\sum\limits_{w^1,w^2}1_{e^2_t(w^2)}(x^2_t)\P(w^1|w^2,x^2_{1:t-1},z_{1:t-1})\P(w^2|x^2_{1:t-1},z_{1:t-1})} \\
 &= \sum_{x^1_t} Q(z_t|x^1_t,x^2_t) \sum_{w^1,w^2} 1_{e^1_t(w^1)}(x^1_t) \nonumber \\
  &\quad\quad \frac{1_{e^2_t(w^2)}(x^2_t)\P(w^1|w^2,z_{1:t-1})\hat{\pi}^2_{t-1}(w^2)}
                   {\sum\limits_{w^1,w^2}1_{e^2_t(w^2)}(x^2_t)\P(w^1|w^2,z_{1:t-1})\hat{\pi}^2_{t-1}(w^2)} \\
   &= \sum_{x^1_t} Q(z_t|x^1_t,x^2_t) \sum_{w^1,w^2} 1_{e^1_t(w^1)}(x^1_t) \nonumber \\
 &\qquad \qquad \frac{1_{e^2_t(w^2)}(x^2_t)\pi_{t-1}(w^1|w^2)\hat{\pi}^2_{t-1}(w^2)}
                     {\sum\limits_{w^1,w^2}1_{e^2_t(w^2)}(x^2_t)\pi_{t-1}(w^1|w^2)\hat{\pi}^2_{t-1}(w^2)} \\
   &= \sum_{x^1_t} Q(z_t|x^1_t,x^2_t) \P(x^1_t|x^2_t,\hat{\pi}^2_{t-1},\pi_{t-1},e_t) \\
   &= \P(z_t|x^2_t,\hat{\pi}^2_{t-1},\pi_{t-1},e_t).
\end{align}
Substituting in the entropy expression we get
\begin{align}
&H(Z_t|X^2_{1:t},Z_{1:t-1}) \nonumber \\
 &= - \sum_{x^2_{1:t-1},z_{1:t-1}}\P(x^2_{1:t-1},z_{1:t-1}) \sum_{x^2_t}\P(x^2_t|\hat{\pi}^2_{t-1},e^2_t) \nonumber \\ &\sum_{z_t}\P(z_t|x^2_t,\hat{\pi}^2_{t-1},\pi_{t-1},e_t) \log \P(z_t|x^2_t,\hat{\pi}^2_{t-1},\pi_{t-1},e_t) \\
 &= \E^{\theta} [ - \sum_{x^2_t}\P(x^2_t|\hat{\pi}^2_{t-1},e^2_t) \sum_{z_t}\P(z_t|x^2_t,\hat{\pi}^2_{t-1},\pi_{t-1},e_t) \nonumber \\
  &\qquad\qquad  \log \P(z_t|x^2_t,\hat{\pi}^2_{t-1},\pi_{t-1},e_t)] \\
  &=\E^{\theta}[h_1(\hat{\Pi}^2_{t-1},\Pi_{t-1},E_t)].
\end{align}
\end{subequations}

The second entropy expression can be written as
\begin{subequations}
\begin{align}
&H(Z_t|X^1_t,X^2_t) \nonumber \\
 &= - \sum_{x^1_t,x^2_t} \P(x^1_t,x^2_t)  \sum_{z_t}Q(z_t|x^1_t,x^2_t) \log Q(z_t|x^1_t,x^2_t) \\
 &= - \int \P(\pi_{t-1},e_t)\sum_{x^1_t,x^2_t} \sum_{w^1,w^2} 1_{e^1_t(w^1)}(x^1_t)1_{e^2_t(w^2)}(x^2_t) \nonumber \\
 &\qquad \pi_{t-1}(w^1,w^2) \sum_{z_t}Q(z_t|x^1_t,x^2_t) \log Q(z_t|x^1_t,x^2_t) \\
 &= \E^{\theta}[ -\sum_{x^1_t,x^2_t} \sum_{w^1,w^2} 1_{e^1_t(w^1)}(x^1_t)1_{e^2_t(w^2)}(x^2_t) \nonumber \\
 &\qquad \pi_{t-1}(w^1,w^2) \sum_{z_t}Q(z_t|x^1_t,x^2_t) \log Q(z_t|x^1_t,x^2_t)] \\
 &=\E^{\theta} [ h_0(\Pi_{t-1},E_t) ].
\end{align}
\end{subequations}

A similar derivation can be followed for the quantity $I(X^2_t;Z_t|X^1_{1:t},Z_{1:t-1})=H(Z_t|X^1_{1:t},Z_{1:t-1})-H(Z_t|X^1_t,X^2_t)$.
For the third quantity $I(X^1_t,X^2_t;Z_t|Z_{1:t-1})=H(Z_t|Z_{1:t-1})-H(Z_t|X^1_t,X^2_t)$ we have
\begin{subequations}
\begin{align}
 &H(Z_t|Z_{1:t-1}) \nonumber \\
 &= - \sum_{z_{1:t-1}}\P(z_{1:t-1}) \sum_{z_t}\P(z_t|z_{1:t-1}) \log \P(z_t|z_{1:t-1}) \\
 &=- \sum_{z_{1:t-1}} \P(z_{1:t-1}) \sum_{z_t} [\sum_{x^1_t,x^2_t}Q(z_t|x^1_t,x^2_t)\P(x^1_t,x^2_t|z_{1:t-1})] \nonumber \\
   &\qquad \qquad \log [\sum_{x^1_t,x^2_t}Q(z_t|x^1_t,x^2_t)\P(x^1_t,x^2_t|z_{1:t-1})],
\end{align}
where
\begin{align}
\P(x^1_t,x^2_t|z_{1:t-1}) &= \sum_{w^1,w^2} 1_{e^1_t(w^1)}(x^1_t)1_{e^2_t(w^2)}(x^2_t)\pi_{t-1}(w^1,w^2) \\
 &= \P(x^1_t,x^2_t|\pi_{t-1},e_t),
\end{align}
and after substituting in the entropy expression we get
\begin{align}
&H(Z_t|Z_{1:t-1}) \nonumber \\
 &=- \sum_{z_{1:t-1}} \P(z_{1:t-1}) \sum_{z_t}
  [\sum_{x^1_t,x^2_t}Q(z_t|x^1_t,x^2_t)\P(x^1_t,x^2_t|\pi_{t-1},e_t)] \nonumber \\
 & \qquad \qquad \log [\sum_{x^1_t,x^2_t}Q(z_t|x^1_t,x^2_t)\P(x^1_t,x^2_t|\pi_{t-1},e_t)] \\
  &=\E^{\theta} [- \sum_{z_t}
  [\sum_{x^1_t,x^2_t}Q(z_t|x^1_t,x^2_t)\P(x^1_t,x^2_t|\pi_{t-1},e_t)] \nonumber \\
 &\qquad \qquad  \log [\sum_{x^1_t,x^2_t}Q(z_t|x^1_t,x^2_t)\P(x^1_t,x^2_t|\pi_{t-1},e_t)]] \\
  &= \E[h_3(\Pi_{t-1},E_t)].
\end{align}
\end{subequations}

Consequently, the mutual information quantities at time $t$ become
\begin{subequations}
\begin{align}
I&(X^1_t;Z_t|X^2_{1:t},Z_{1:t-1}) \nonumber \\
 &= \E^{\theta}[ h_1(\hat{\Pi}^2_{t-1},\Pi_{t-1},E_t) -h_0(\Pi_{t-1},E_t) ] \\
 &= \E^{\theta}[ i_1(\hat{\Pi}^2_{t-1},\Pi_{t-1},E_t) ] \\
I&(X^2_t;Z_t|X^1_{1:t},Z_{1:t-1}) \nonumber \\
 &= \E^{\theta}[ h_2(\hat{\Pi}^1_{t-1},\Pi_{t-1},E_t) -h_0(\Pi_{t-1},E_t) ] \\
 &= \E^{\theta}[ i_2(\hat{\Pi}^1_{t-1},\Pi_{t-1},E_t) ] \\
I&(X^1_t,X^2_t;Z_t|Z_{1:t-1}) \nonumber \\
 &= \E^{\theta}[ h_3(\Pi_{t-1},E_t) -h_0(\Pi_{t-1},E_t) ]\\
 &= \E^{\theta}[ i_3(\Pi_{t-1},E_t) ]
\end{align}
\end{subequations}

}
\end{IEEEproof}

We remark at this point that the presence of the new quantity $\hat{\pi}^i_t$ is surprising and requires further investigation since it does not appear in the DSAHT formulation of Section~\ref{sec:dsaht}.
It is a marginal posterior of each message conditioned on the transmitted signal of the corresponding user and the received signal and we refer to it as the \emph{private state} of user $i$. This new quantity gives further insight as to  why the problem of finding the MAC feedback capacity has resisted solution till now.
To see why, take for instance the instantaneous quantity $H(Z_t|X^2_{1:t},Z_{1:t-1})$ related to~\eqref{eq:th2a}.
This quantity depends on the distributions
$\P(x^2_t|x^2_{1:t-1},z_{1:t-1})=q^2_t(x^2_t|x^2_{1:t-1},z_{1:t-1})$, but it also depends on a distribution of the form $\P(x^1_t|x^2_{1:t},z_{1:t-1})$ which is not a simple function of $q^i_t$, but it depends on the entire sequence of $(q^1_{\tau},q^2_{\tau})_{\tau\in\{1,\ldots,t\}}$.
It is exactly this long-range dependence of the instantaneous quantities
at time $t$ on all previous distributions up to time $t$ that makes this optimization problem unwieldy.
\optv{arxiv} {
The methodology we are following allows us to
(a) simplify the structure of the distributions $q^i_t$, e.g., $q^2_t(x^2_t|x^2_{1:t-1},z_{1:t-1})=\P(x^2_t|\hat{\pi}^2_{t-1},e^2_t)=\sum_{w^2} 1_{e^2_t(w^2)}(x^2_t)\hat{\pi}^2_{t-1}(w^2)$ and more importantly (b)
to simplify expressions such as  $\P(x^1_t|x^2_{1:t},z_{1:t-1})=\P(x^1_t|x^2_t,\hat{\pi}^2_{t-1},\pi_{t-1},e_t)$.
} 

We now comment on the significance of this theorem.
Fix $\underline{\lambda}\in\Real^3_+$.
Theorem~\ref{th:capacity} shows that the expression $I_n(\underline{\lambda})$ in~\eqref{eq:region_to_cost_fb} involved in evaluating the channel capacity
can be expressed as
\begin{equation}
I_n(\underline{\lambda})=\frac{1}{n} \sum_{t=1}^n \E^{\theta}[i(\Pi_{t-1},\hat{\Pi}_{t-1},E_t;\underline{\lambda})].
\end{equation}
Furthermore, the unstructured optimization problem for finding $C_n(\underline{\lambda})$ in~\eqref{eq:region_to_cost_fb}
can now be restated as
\begin{equation}
C_n(\underline{\lambda}) = \sup_{\theta} \frac{1}{n} \sum_{t=1}^n \E^{\theta}[i(\Pi_{t-1},\hat{\Pi}_{t-1},E_t;\underline{\lambda})].
\end{equation}
The above expression hints at thinking of the quantity $C_n(\underline{\lambda})$ as the average reward received
from a dynamical system with ``state'' $(\hat{\Pi}_{t-1},\Pi_{t-1})$ partially controlled by the
encoding functions $E_t=\theta_t[\Pi_{t-1}]$, and optimized over all such policies.
What remains to show is that indeed the pair $(\hat{\Pi}_{t-1},\Pi_{t-1})$ is the state of a controlled dynamical system.
The result is stated in the following theorem.
\begin{theorem}
$(\hat{\Pi}_{t-1},\Pi_{t-1})_{t\geq 1}$ is a Markov process controlled by the quantity $E_t$, i.e.,
\begin{align}
\P^{\theta}(\hat{\pi}_{t},\pi_{t}|\hat{\pi}_{1:t-1},\pi_{1:t-1},e_{1:t})
 = \P(\hat{\pi}_{t},\pi_{t}|\hat{\pi}_{t-1},\pi_{t-1},e_{t}),
\end{align}
and the latter distribution does not depend on the policy $\theta$.
\end{theorem}
\begin{IEEEproof}
\optv{2col}{
The proof is omitted due to space limitations. It can be found in the full version of the paper in~\cite{AnPr20arxiv}.
}
\optv{arxiv}{
We have
\begin{subequations}
\begin{align}
&\P(\hat{\pi}_{t},\pi_{t}|\hat{\pi}_{1:t-1},\pi_{1:t-1},e_{1:t}) \nonumber \\
 &=\sum_{z_t,x^1_t,x^2_t}
  1_{F(\pi_{t-1},e_t,z_t)}(\pi_t)
  1_{\hat{F}(\hat{\pi}^1_{t-1},e^1_t,x^1_t)}(\hat{\pi}^1_t)
  1_{\hat{F}(\hat{\pi}^2_{t-1},e^2_t,x^2_t)}(\hat{\pi}^2_t) \nonumber \\
 &\qquad\qquad Q(z_t|x^1_t,x^2_t)\sum_{w^1,w^2}1_{e^1_t(w^1)}(x^1_t)1_{e^2_t(w^2)}(x^2_t) \nonumber \\
 &\qquad\qquad\qquad \P^{\theta}(w^1,w^2|\hat{\pi}_{1:t-1},\pi_{1:t-1},e_{1:t})\\
 &=\sum_{z_t,x^1_t,x^2_t}
  1_{F(\pi_{t-1},e_t,z_t)}(\pi_t)
  1_{\hat{F}(\hat{\pi}_{t-1},e^1_t,x^1_t)}(\hat{\pi}^1_t)
  1_{\hat{F}(\hat{\pi}_{t-1},e^2_t,x^2_t)}(\hat{\pi}^2_t)   \nonumber \\
 &\qquad\qquad Q(z_t|x^1_t,x^2_t)\sum_{w^1,w^2}1_{e^1_t(w^1)}(x^1_t)1_{e^2_t(w^2)}(x^2_t)\nonumber \\
 &\qquad\qquad\qquad \hat{\pi}^1_{t-1}(w^1)\hat{\pi}^2_{t-1}(w^2) \\
 &=\P(\hat{\pi}_{t},\pi_{t}|\hat{\pi}_{t-1},\pi_{t-1},e_{t}),
\end{align}
where we have used the fact that
\begin{align}
&\P(w^1,w^2|\hat{\pi}_{1:t-1},\pi_{1:t-1},e_{1:t}) \nonumber \\
 &= \hspace*{-0.5cm}\sum_{\substack{x^1_{1:t-1},x^2_{1:t-1},z_{1:t-1}:\\\hat{\pi}_{1:t-1},\pi_{1:t-1},e_{1:t}}} \hspace*{-1cm}\P(w^1,w^2|x^1_{1:t-1},x^2_{1:t-1},z_{1:t-1},\hat{\pi}_{1:t-1},\pi_{1:t-1},e_{1:t}) \nonumber \\
  &\qquad\qquad  \P(x^1_{1:t-1},x^2_{1:t-1},z_{1:t-1}|\hat{\pi}_{1:t-1},\pi_{1:t-1},e_{1:t}) \\
  &= \hat{\pi}^1_{t-1}(w^1)\hat{\pi}^2_{t-1}(w^2) \nonumber \\
  &\sum_{\substack{x^1_{1:t-1},x^2_{1:t-1},z_{1:t-1}:\\\hat{\pi}_{1:t-1},\pi_{1:t-1},e_{1:t}}}
   \hspace*{-1cm}\P(x^1_{1:t-1},x^2_{1:t-1},z_{1:t-1}|\hat{\pi}_{1:t-1},\pi_{1:t-1},e_{1:t}) \\
 &= \hat{\pi}^1_{t-1}(w^1)\hat{\pi}^2_{t-1}(w^2).
\end{align}
\end{subequations}
}
\end{IEEEproof}


%
%

\clearpage

\optv{2col}{
\input{isit20_final.bbl}
}
\optv{arxiv}{

}
\end{document}